# General time-reversal equivariant neural network potential for magnetic materials

Hongyu Yu[a,b,*], Boyu Liu[a,b,*], Yang Zhong[a,b], Liangliang Hong[a,b], Junyi Ji[a,b], Changsong Xu[a,b], Xingao Gong[a,b], Hongjun Xiang[a,b,†]

[a]*Key Laboratory of Computational Physical Sciences (Ministry of Education), Institute of Computational Physical Sciences, State Key Laboratory of Surface Physics, and Department of Physics, Fudan University, Shanghai 200433, China*

[b]*Shanghai Qi Zhi Institute, Shanghai 200030, China*

*These authors contributed equally

[†]Email: hxiang@fudan.edu.cn

## Abstract

This study introduces time-reversal E(3)-equivariant neural network and SpinGNN++ framework for constructing a comprehensive interatomic potential for magnetic systems, encompassing spin-orbit coupling and noncollinear magnetic moments. SpinGNN++ integrates multitask spin equivariant neural network with explicit spin-lattice terms, including Heisenberg, Dzyaloshinskii-Moriya, Kitaev, single-ion anisotropy, and biquadratic interactions, and employs time-reversal equivariant neural network to learn high-order spin-lattice interactions using time-reversal E(3)-equivariant convolutions. To validate SpinGNN++, a complex magnetic model dataset is introduced as a benchmark and employed to demonstrate its capabilities. SpinGNN++ provides accurate descriptions of the complex spin-lattice coupling in monolayer $CrI_3$ and $CrTe_2$, achieving sub-meV errors. Importantly, it facilitates large-scale parallel spin-lattice dynamics, thereby enabling the exploration of associated properties, including the magnetic ground state and phase transition. Remarkably, SpinGNN++ identifies a new ferrimagnetic state as the ground magnetic state for monolayer $CrTe_2$, thereby enriching its phase diagram and providing deeper insights into the distinct magnetic signals observed in various experiments.

# Introduction

In recent years, machine learning (ML) has found widespread use in material modeling due to its ability to strike a balance between speed and accuracy[1–3]. ML considerably reduces the computational cost compared to ab initio calculation, enabling the feasible simulation of large systems[4,5]. Many ML models use the descriptors of materials as input and learn the corresponding mapping function between the input descriptor and the material's energy[6–9]. The graph neural network (GNN) has emerged as a leading technique for studying molecules and materials, specifically with the development of molecular and crystal GNNs, which serve as end-to-end natural descriptors for materials[10–17]. The preservation of E(3) symmetry in GNN graph representations, including translation, rotation, parity and permutation, occurs naturally. The recent state-of-the-art developments in GNN include the E(3)-equivariant graph neural network (ENN)[18–21] (e.g., NequIP[22] and Allegro[5]). Several benchmark tests performed on small molecular systems and bulk systems have demonstrated that ENN can achieve comparable accuracy to that of invariant GNN with significantly fewer training points[22,23].

The magnetic potential plays a crucial role in magnetic systems, particularly in determining the magnetic ground state, facilitating spin-lattice dynamics, and conducting Monte Carlo simulations. Unlike the usual atomic potentials, which do not consider the degrees of freedom of spin, the magnetic potential must achieve an accuracy within sub-meV per atom due to the small energy fluctuations among different spin configurations. Some machine-learning potentials have been devised for magnetic materials[24–31]. Yu et al. developed SpinGNN to investigate the magnetic phase transition of $BiFeO_3$[25] and Matteo et al. augmented magnetic ACE to investigate the magnetic behavior of iron[31]. However, an general accurate spin-lattice neural network potential method that incorporates the effect of spin-orbit coupling is still lacking.

In this work, we introduce SpinGNN++, a comprehensive equivariant magnetic potential framework that integrates MSENN (multitask spin equivariant neural network) and TENN (time-reversal equivariant neural network) to incorporate the spin-orbit effect and complex spin-lattice interactions, thus enabling generalization to most magnets. We have developed Spin-Allegro++ in this study, exhibiting high precision and efficiency in capturing intricate spin-lattice interactions in magnetic materials, as demonstrated by a complex magnetic model dataset, monolayer $CrI_3$, and $CrTe_2$. Our experiments demonstrate that Spin-Allegro++ can reproduce the magnetic phase transition in 2D ferromagnetic $CrI_3$. Interestingly, we reveal a new ground magnetic state for monolayer $CrTe_2$.

## Results

### The SpinGNN++ framework

The SpinGNN++ framework, as depicted in Fig. 1 (a)(b) is an extension of the original SpinGNN framework[25] to account for spin-orbit coupling effects and incorporate more complex spin-lattice interactions. SpinGNN++ comprises two distinct neural network architectures: MSENN and TENN. MSENN employs updated scalar and tensor features on edges in ENN to treat explicitly the Heisenberg, Dzyaloshinskii–Moriya (DM), anisotropic symmetric exchange, single ion anisotropy (SIA), and biquadratic spin-lattice interactions between magnetic atoms. Simultaneously, TENN is designed to learn complex high-order non-linear spin-lattice interactions. The total energy for SpinGNN++ is represented as:

$$E^{\text{total}} = E^{\text{MSENN}} + E^{\text{TENN}} \qquad (1)$$

The spin force (denoted as $\vec{\omega_i} = -\partial E/\partial \vec{s_i}$), which is defined as the negative gradient of the total energy with respect to the spin vector of atom $i$, can be computed using auto-differentiation, along with atomic force and stress. In this article, the input vector $\vec{s}$ in MSENN and TENN is a unit vector. If the magnitude of the magnetic moment varies with the atomic structure and spin configuration, a distinct multilayer perceptron (MLP) within the output of TENN can be used for prediction. Conversely, if the magnitude of the magnetic moment remains constant across different atomic structures and spin configurations, as seen in most magnetic materials[32,33], it can be excluded from the prediction and treated as a constant during spin-lattice dynamics.

***Multitask Spin Equivariant Neural Network (MSENN).*** We first introduce the framework of MSENN [see Fig. 1(a)]. MSENN is based on ENN and the energy is represented as:

$$E^{\text{MSENN}}(\vec{r},\vec{s}) = E_r(\vec{r}) + E_J(\vec{r},\vec{s}) + E_{SIA}(\vec{r},\vec{s}) + E_{Biquad}(\vec{r},\vec{s})$$

$$= \sum_i h_i^{\text{MSENN}}(\vec{r}) + \sum_{ij}\sum_{\alpha,\beta} \mathbf{J}_{\alpha,\beta}^{ij}(\vec{r})s_{i\alpha}s_{j\beta} + \sum_i \sum_{\alpha,\beta} \mathbf{A}_{\alpha,\beta}^{i}(\vec{r})s_{i\alpha}s_{i\beta}$$

$$+ \sum_{ij} K_{ij}(\vec{r})(\vec{s}_i \cdot \vec{s}_j)^2 \quad (2)$$

where $E_r$ represents the potential energy only related to the atomic structure, $E_J$ describes the general second-order magnetic pair interactions (including the Heisenberg, DM, and anisotropic symmetric exchange interactions), $E_{SIA}$ denotes the SIA energy, $E_{Biquad}$ is the biquadratic interactions interaction, $h_i^{\text{MSENN}}(\vec{r})$ is the structure-related component of the local atomic energy of atom *i*, $\mathbf{J}_{\alpha,\beta}^{ij}(\vec{r})$ and $K^{ij}(\vec{r})$ represent the 3 × 3 J matrix and biquadratic coefficient of the bond between magnetic ions *i* and *j*, respectively, and $\mathbf{A}_{\alpha,\beta}^{i}(\vec{r})$ denotes the SIA matrix of the atom *i*. MSENN takes the same inputs as ENN, consisting of information about the structure and atomic species. The ENN uses high-order tensor features to predict the J and SIA matrices with equivariant layers, while the scalar features are used to predict the biquadratic coefficients with MLP. In the case of neglecting SOC, MSENN solely predicts Heisenberg $J_{ij}(\vec{r})$ and the biquadratic coefficients with scalar features with two separate MLPs. These terms are dependent solely on the local atomic position environment and are independent of spin configuration. MSENN not only includes the most dominant and common explicit spin-lattice potential but also provides the exchange parameters with a physical insight, which is not provided in most machine-learning magnetic potential.

*Time-reversal equivariant neural networks (TENN).* Although the dominant spin interactions are taken into account in MSENN, many-body (e.g., four-body ring-exchange[34]) may play a crucial role in some magnets. Here, we design the TENN framework to consider many-body and high-order spin interactions. The TENN magnetic potential framework is introduced as demonstrated in Fig. 1(b) and the energy is represented as:

$$E^{\text{TENN}}(\vec{r},\vec{s}) = \sum_i h_i^{\text{TENN}}(\vec{r},\vec{s}) \qquad (8)$$

where $h_i^{\text{TENN}}(\vec{r},\vec{s})$ represents the local atomic energy of atom $i$ in relation to the position and spin vector of the surrounding atoms. The inputs of TENN not only include atomic displacement vector $\vec{r_{ij}}$, atomic number $z_i$ in ENN, but also the spin vector $\vec{s_i}$ (a pseudo-vector) of magnetic atoms. Note that the spin vector $\vec{s}$ transforms differently under spatial inversion **P** and time-reversal **T** from the usual atomic displacement vector $\vec{r}$ ($\mathbf{P}\vec{r}=-\vec{r}$, $\mathbf{T}\vec{r}=\vec{r}$, $\mathbf{P}\vec{s}=\vec{s}$, $\mathbf{T}\vec{s}=-\vec{s}$). As the energy should remain the same under spatial inversion **P** and time-reversal **T** operations, one can easily show that $E(\vec{r},\vec{s}) = E(\mathbf{P}\vec{r},\mathbf{P}\vec{s}) = E(-\vec{r},\vec{s})$ and $E(\vec{r},\vec{s}) = E(\mathbf{T}\vec{r},\mathbf{T}\vec{s}) = E(\vec{r},-\vec{s})$. Since the usual ENN can only deal with the usual vector $\vec{r}$, if one directly adopts ENN to describe $E(\vec{r},\vec{s})$, these energy relations will not be satisfied. The purpose of our developed TENN is to treat the spin vectors properly.

Then, we present the fundamental concept of TENN based on extended irreducible representations (often abbreviated as irreps) with an additional time-reversal index derived from the irreps of ENN[35]. The application of E(3)-equivariant representations and operations with geometric tensors is a critical feature of ENN. Essentially, the ENN framework utilizes high-order tensor products based on the irreps of the $O(3)$ group that are indexed by the index of a rotation and a parity order. Furthermore, physical quantities can be categorized into two classes based on the response under the time-reversal operation. Quantities that remain unchanged are identified by an even time-reversal index, $t = 1$, whereas those that change sign are represented by an odd time-reversal index, $t = -1$, such as the spin and velocity of an atom. Each variable, including scalars, vectors, and higher-order tensors, in TENN, is assigned a time-reversal index. When two variables are added or subtracted, their time-reversal indices are verified and modified. Only tensors with the same time-reversal index could be added or subtraction. Similarly, when two tensors **x** and **y** with $t_1$ and $t_2$ as time-reversal indices are multiplied, the resulting tensor inhabits an irrep indexed by the product of both time-reversal indices, $t_{out} = t_1 t_2$. Consequently, the tensor in TENN is specified by three orders: the rotation order $\ell \geq 0$, parity order $p \in \{-1, 1\}$ and the time reversal order $t \in \{-1, 1\}$. The core operation of TENN is the tensor product of representations, which combines two tensors **x** and **y** with irrep $\ell_1$, $p_1$, $t_1$ and $\ell_2$, $p_2$, $t_2$ yielding an output that inhabits an irrep $\ell_{out}$, $p_{out}$, $t_{out}$ that satisfies particular conditions, including $|\ell_1 - \ell_2| \leq \ell_{out} \leq |\ell_1 + \ell_2|$, $p_{out} = p_1 p_2$ and $t_{out} = t_1 t_2$

$$(\mathbf{x} \otimes \mathbf{y})_{\ell_{out}, m_{out}} = \sum_{m_1, m_2} \begin{pmatrix} \ell_1 & \ell_2 & \ell_{out} \\ m_1 & m_2 & m_{out} \end{pmatrix} \mathbf{x}_{l_1, m_1} \mathbf{y}_{l_2 m_2} \quad (7)$$

where $\begin{pmatrix} \ell_1 & \ell_2 & \ell_{out} \\ m_1 & m_2 & m_{out} \end{pmatrix}$ is the Wigner $3j$ symbol. By including an additional representation index for time-reversal symmetry, we can utilize existing equivariant models (such as NequIP[22], and Allegro[5]) to construct the time-reversal equivariant model. To build the time-reversal equivariant neural network with a higher level of symmetry, we extend the popular e3nn package[35] to the T-e3nn package.

**Spin-Allegro++.** In this work, we selected Allegro[5] as a SpinGNN++ instance. Allegro is a state-of-the-art ENN that exhibits high accuracy and a superior level of transferability to out-of-distribution data. Moreover, Allegro can enable large-scale simulations involving hundreds of millions of atoms that can be parallelized across devices, thanks to its localization in representation. Spin-Allegro++ inherits these aforementioned advantages as the SpinGNN++ version of Allegro. We employed

Spin-Allegro++ for the commonly occurring noncollinear magnetic moment scenario, accounting for SOC effects, as shown in Fig. 2 (see details of the model in SM).

In Spin-Allegro++, MSENN facilitates the exploration of explicit bilinear and biquadratic magnetic potentials, while TENN is adopted at modeling complex high-order magnetic potentials, thereby accommodating a broad spectrum of spin-lattice interactions owing to its strong expressiveness. Both models can be employed independently or collaboratively to establish the magnetic potential of an ensemble based on specific requirements. When SOC is negligible, MSENN exclusively predicts Heisenberg and biquadratic interactions, while TENN performs convolutions solely on the spin vector. Consequently, the choice of the model for creating the magnetic potential landscape depends on the adaptability and is determined by the type of magnetic interactions prevalent in the material of interest.

**Tests on Spin-Allegro++**

*Artificial high-order spin-lattice model dataset with spin-orbit couplings*. Utilizing an artificially constructed high-order spin-lattice model, a magnetic potential dataset are generated and employed to demonstrate the capability and accuracy of Spin-Allegro++. Additionally, the dataset is provided as benchmark data for magnetic neural network potential, considering noncollinear magnetic moment and SOC. The dataset model is built to describe the spin-lattice interactions in MnGe-type[36] structure showing in Fig. 3(a), encompassing 4149 terms of spin-lattice interactions from one-body one-order up to three-body four-order forms. These include common interactions such as Heisenberg, DM, Kitaev, biquadratic, and SIA interactions, as well as other complex interactions with the interacting coefficients varying from 0 to 1 eV. Random displacements and spin vectors are imposed on the structures, and relative energy is calculated based on the artificial model using PASP[37]. A total of 7120 data points are generated and subsequently divided into training (5696 samples), validation (712 samples), and testing (712 samples) sets. The target variable is energy only. Spin-Allegro++ is employed with two layers of MSENN and TENN to fit the dataset, achieving a low MAE (mean absolute error), and an R2 score of 0.9979 on the test dataset, as shown in Fig. 3(b). Consequently, we infer that Spin-Allegro++ characterizes complex high-order spin-lattice interactions with a high degree of accuracy.

***Monolayer $CrI_3$***. Two-dimensional (2D) magnetic materials play a crucial role in the advancement of spin electronics. Among these materials, monolayer chromium triiodide ($CrI_3$) is particularly notable. This 2D material consists of a single layer of $CrI_3$ crystals and has been successfully synthesized and characterized at low temperatures[38]. The monolayer exhibits a 1.2 eV bandgap and significant magnetic anisotropy[39], primarily attributed to spin-orbit coupling (SOC) effects. The pronounced magnetic anisotropy in the out-of-plane direction results in long-range ferromagnetic ordering in the $CrI_3$ monolayer, contributing to its stable out-of-plane ferromagnetism and a Curie temperature of 45 K[40].

A dataset containing 6122 data points is generated through first-principles calculations on monolayer $CrI_3$ and divided into training (4900 samples), validation (611 samples), and testing (611 samples) sets. The radial cutoff of the networks is set as 8.0 Å to encompass atoms up to the third magnetic neighbor. Multiple Spin-Allegro++ networks, each considering different interactions, are employed for training and evaluation on the test dataset, with the results presented in Table 1 and the details available in supplemental materials (SM). The disparity in accuracy among models with different terms leads to the conclusion that the J matrix encompassing Heisenberg, DM and Kitaev interactions[41,42] predominates in monolayer $CrI_3$ with subtle and neglectable SIA and high-order spin interaction. Further training solely on energy yields an accuracy of 0.110 meV/atom for the MSENN with J matrix model, which closely approaches that of the MSENN & TENN model. This model accurately describes the energy landscape and is subsequently utilized for additional experiments.

| Spin-Allegro++ Models | Energy (meV/atom) | Force (meV/Å) | Stress (meV/Å$^3$) |
|---|---|---|---|
| MSENN & TENN | 0.101 | 6.00 | 0.111 |
| MSENN with J matrix, Biquadratic and SIA | 0.173 | 7.79 | 0.116 |
| MSENN with J matrix and SIA | 0.169 | 7.88 | 0.116 |
| MSENN with J matrix | 0.153 | 8.35 | 0.123 |
| Allegro w/o spin | 2.463 | 25.76 | 0.169 |

Table 1. Results comparison of Spin-Allegro++ models with different configurations on monolayer CrI$_3$ dataset, measured by mean absolute error.

To assess the potential accuracy, we relax the atomic magnetic structure including the degrees of freedom for the system cell, atomic positions, and spin configuration utilizing the conjugate gradient (CG)[43] and annealing[44] algorithm to explore the magnetic ground state of CrI$_3$. This process begins with a 3×3×1 supercell consisting of 72 atoms and 18 Cr atoms with random spin configurations, and culminates in an out-of-plane ferromagnetic state, shown in Fig. 4(a), consistent with experiments findings[38]. The final magnetic structure is validated through high-accuracy density functional theory (DFT) calculations with average and maximum atomic force of 7.90 meV/Å and 22.5 meV/Å, respectively demonstrating the effectiveness of the relaxation. Furthermore, the energy disparities between different magnetic states — the out-of-plane ferromagnetism (OP-FM), the in-plane ferromagnetism (IP-FM), the out-of-plane antiferromagnetism (OP-AFM), and the in-plane antiferromagnetism (IP-AFM)—are calculated with DFT and Spin-Allegro++, as shown in Table 2. The subtle energy differences and energy rankings are accurately reproduced, facilitating the precise prediction of the magnetic ground state and the accuracy of Spin-Allegro++.

| Method | Magnetic state | OP-FM | IP-FM | OP-AFM | IP-AFM |
|---|---|---|---|---|---|
| DFT | | 0.00 | 0.142 | 7.26 | 7.07 |
| Spin-Allegro++ | | 0.00 | 0.267 | 6.91 | 6.55 |

Table 2. The energy difference per atom between different magnetic states of monolayer $CrI_3$ is calculated using first-principle calculations and Spin-Allegro++ and expressed in units of meV/atom.

Subsequently, we perform spin-lattice simulations[33,45] across a range of temperatures to investigate the magnetic phase transition in monolayer $CrI_3$. The ferromagnetic vector of the system is tracked from 1 K to 100 K, as illustrated in Fig. 4(c). The findings reveal that as the temperature rises, the ferromagnetic vector diminishes, indicating a magnetic transition from the out-of-plane ferromagnetic to the paramagnetic state, with the estimated Curie temperature being 70 K.

**Application: Monolayer $CrTe_2$**

$CrTe_2$ is an intriguing material due to its ability to maintain long-range magnetic order at room temperature, even with fewer layers[46,47]. Specifically, monolayer $CrTe_2$ exhibits a diverse array of magnetic ground states under strong spin-lattice coupling, presenting the potential for mutually regulating structure and magnetic state. However, the ground-state magnetic order of monolayer $CrTe_2$ is under hot debate. Experimental studies have reported ferromagnetic order in ultrathin $CrTe_2$[46,47], whereas a recent study has identified evidence of a zigzag (ZZ) antiferromagnetic (AFM) state in monolayer $CrTe_2$[48]. Ongoing debates centered on employing first-principle calculations to ascertain the lowest energy lattice structure under a fixed magnetic configuration, encompassing the frustrated triangle antiferromagnetic, ferromagnetic (FM), ZZ-AFM, ABAB, and AABB striped AFM (sAFM)[49–52].

A dataset for monolayer CrTe$_2$, consisting of 7077 data points, is generated using first-principles calculations and divided into training (5661 samples), validation (708 samples), and testing (708 samples) sets (see details in SM). The radial cutoff of networks is set at 7.5 Å to incorporate atoms up to the third magnetic neighbor. Various networks, each incorporating different numbers of energy terms, are trained, with the results presented in Table 3. It is demonstrated that the model with SIA and biquadratic terms enhances the accuracy of energy prediction by 0.60 and 0.27 meV/Cr, respectively. Furthermore, a significant improvement of 1.809 meV/Cr with TENN-Allegro implies that complex implicit many-body high-order spin-lattice interactions exert a substantial influence in this system, going beyond the commonly investigated Heisenberg, DM, Kitaev, SIA, and biquadratic spin-lattice interactions. For an explicit spin-lattice model of monolayer CrTe2, it is essential to consider complex many-body high-order terms.

| Models | Energy (meV/atom) | Force (meV/Å) | Stress (meV/Å$^3$) |
|---|---|---|---|
| MSENN & TENN | 0.637 | 18.8 | 0.223 |
| MSENN with J matrix, Biquadratic and SIA | 1.24 | 20.7 | 0.239 |
| MSENN with J matrix and SIA | 1.35 | 22.3 | 0.230 |
| MSENN with J matrix | 1.55 | 21.4 | 0.239 |
| Allegro without spin | 6.06 | 65.4 | 0.359 |

Table 3. Assessment of the performance of different Spin-Allegro++ models on the monolayer CrTe$_2$ dataset, as measured by mean absolute error.

The complex magnetic ground state of monolayer CrTe$_2$ is investigated using trained models with the same optimization procedures as for CrI$_3$. The relaxation commences with an 6×4$\sqrt{3}$×1 supercell, which is large enough to accommodate most magnetic states yet computationally feasible for the first-principle calculations required for validation. This supercell comprises 144 atoms, with 48 Cr atoms in a random spin configuration. Employing the optimal Spin-Allegro++ model, encompassing MSENN and TENN, reveals a ferrimagnetic (Ferri) state as the ground magnetic state with 1/3 of the Cr atoms exhibiting an "up" orientation surrounded by 2/3 displaying a "down" orientation, depicted in Fig. 4(b).

Previous research has demonstrated that monolayer $CrTe_2$ can endure a significant range of strain with only a slight alteration in the Cr-Te bond within the range of -10% to +10%[49]. This characteristic lends significance to the examination of the strain-induced magnetic phase diagram in a system with strong spin-lattice coupling. As Spin-Allegro++ reveals a previously unexplored potential magnetic ground state, we revisit the magnetic phase diagram under epitaxy strain with the introduction of a Ferri state. Initially, a $6\times2\sqrt{3}\times1$ supercell is applied to accommodate 5 magnetic states. It is worth noting that the commonly used $2\times2\sqrt{3}\times1$ cell, as depicted in the green rectangle of Fig. 5(a), is incapable of constructing the Ferri state under periodic boundary conditions. This limitation also extends to smaller cells in some other studies, rendering the Ferri state unattainable in the examination of magnetic states in monolayer $CrTe_2$. Subsequently, energies for different magnetic states are computed under varied strains ranging from -8% to 8%, individually applied to the a and b axes. The data points are calculated at every 2% strain, resulting in a $9\times9$ data mesh in the phase diagram.

The strain-induced magnetic phase diagram, presented in Fig. 6, reveals that the Ferri state occupies a substantial region under compressive strain, contradicting earlier assumptions of a zigzag magnetic ground state in previous works[52,53]. Meanwhile, the regions corresponding to FM AABB and zigzag align with the findings from prior literature. To incorporate the SOC effect, the phase diagram is further recalculated. Due to the intricate magnetic interactions in $CrTe_2$, the easy axis varies across different magnetic configurations. Consequently, the energies of distinct magnetic states are computed with spin orientations along their respective easy axes (see details in SM), and the phase boundary with SOC is delineated by the red dashed line. Additionally, the freestanding structures of Ferri, ZZ, AABB and FM with fully relaxed lattice and atom positions, are indicated by yellow symbols. Remarkably, when taking the spin-orbit coupling (SOC) effect into account, the energy of the freestanding Ferri state is found to be 3.82 meV/atom lower than ZZ, 6.63 meV/atom lower than AABB and 20.11 meV/atom lower than FM, signifying the Ferri state as the global magnetic ground state. As revealed by the lattice-magnetic phase diagram, the substrate lattice may contribute to the divergent observations of antiferromagnetic[48] and ferromagnetic [49] signals in distinct experiments. The identification of the ferrimagnetic state in monolayer $CrTe_2$ through SpinGNN++ has enhanced the understanding of the material's magnetic phase diagram and sheds light on the disparate results of antiferromagnetic and ferromagnetic signals observed in various experiments. By adjusting the strain within the largely unexplored region of high compressive strain, it may be feasible to produce monolayer ferrimagnets that hold substantial practical value. Considering the small energy gap between Ferri and ZZ and the phase diagram on lattice, it is feasible to tune the magnetic state by manipulating the substrate lattice, thereby transitioning between antiferromagnetic and ferromagnetic states.

Finally we employ spin-lattice simulation[33,45] to estimate the Curie temperature of the Ferri magnetic state in monolayer $CrTe_2$. Different from the usual simulation where only the spin degrees of freedom are taken into account, here both atomic displacement and spin degrees of freedom are considered. The net magnetic moment of the system is monitored from 0.1 K to 120.1 K, as illustrated in Fig. 4(d). Notably, at 0.1 K, the average magnetic moment of a Cr atom is approximately 1 $\mu_B$, with 2/3 of the Cr atoms exhibiting an "up" orientation and 1/3 displaying a "down" orientation characteristic of the Ferri magnetic state. With increasing temperature, the ferromagnetic vector diminishes from a value near one to zero, indicative of a magnetic transition from the Ferri to the paramagnetic state. The transition temperature is estimated to be 60 K.

# Discussion

We introduce a comprehensive framework for deep-learning magnetic potential that combines MSENN and TENN to achieve high accuracy in predicting energies, forces, and stresses associated with spin-lattice interactions. Our SpinGNN++ method integrates both structural and magnetic degrees of freedom, accommodating noncollinear magnetic moments, and incorporating spin-orbit coupling in explicit Heisenberg, DM, Kitaev, biquadratic, and implicit high-order spin-lattice interactions. TENN is used to model complex high-order magnetic potentials, while MSENN is employed to predict explicit and general magnetic interactions. SpinGNN++ has demonstrated high efficiency and accuracy for large-scale spin-lattice simulations, as seen in monolayer $CrI_3$ and $CrTe_2$, achieving sub-meV per atom accuracy. With SpinGNN++, we discover a new ferrimagnetic ground state in monolayer $CrTe_2$, which may resolve the disputes on the magnetic ground state in monolayer $CrTe_2$. Our work suggests that SpinGNN++ can be adopted to predict the ground state, thermodynamic and kinetic properties of complex large scale magnetic systems with nearly first-principles fidelity.

In comparison to SpinGNN[25], SpinGNN++ can additionally account for spin-orbit couplings and more complex related spin-lattice interactions using time-reversal E(3) equivariant neural networks. This is particularly significant for various magnetic systems, especially 2D magnets. Our framework enables the investigation of magnetic systems that were previously inaccessible with time-reversal equivariant magnetic potential. The SpinGNN++ framework can be implemented with most equivariant neural network architectures, such as MACE[23] and NequIP[22] to be implemented just as Allegro[5], and can be utilized to develop the magnetic potential of an ensemble based on the specific situation. With the development of the equivariant neural network potential, it is expected that more powerful spin-lattice interactions can be built with the SpinGNN++ framework, such as Allegro-v2[54] which is recently proposed and offers faster implementation. Additionally, MSENN can accommodate more explicit spin-lattice interactions in MSENN in either tensor or scalar forms, analogous to the J matrix and biquadratic interactions. Considering the generalization of the SpinGNN++ framework, it is expected to find widespread use in the investigation of magnetic materials.

The T-e3nn package is developed to elevate the E(3) symmetry of the equivariant model to a higher level of time-reversal E(3) symmetry based on the widely-used e3nn package[35], allowing for the complete representation of all physical quantities. Any E(3) equivariant model can be easily transformed into a time-reversal E(3) equivariant model using the T-e3nn package. Furthermore, the phase space, including atomic position and velocity $\{\vec{r_{ij}}, \vec{v_i}\}$, can be accommodated with the consideration of time reversal symmetry, which may find utility in the analysis of molecular trajectories.

# Methods

## Software

The time reversal Euclidean neural networks package (T-e3nn) is an extension of the e3nn package with version 0.5.1[35] and codes are available at https://github.com/Hongyu-yu/T-e3nn. SpinGNN++ is developed based on the modified NequIP package with version 0.5.6 available at https://github.com/mir-group/nequip, Allegro package with version 0.2.0 available at https://github.com/mir-group/allegro, as well as Pytorch[55] with version 1.11.0 and Python with 3.9.13. The LAMMPS is built based on the LAMMPS code available at https://github.com/lammps/lammps.git under git commit 9b989b186026c6fe9da354c79cc9b4e152ab03af with the pair_allegro code available at https://github.com/mir-group/pair_allegro. The VESTA[56] software is used to generate figures of crystal structures. Matplotlib[57] is used for plotting results.

## Spin-lattice dynamics simulations

We performed spin-lattice dynamics simulations using the SPIN package[58] in LAMMPS[59]. To accomplish accurate integration of the Landau-Lifshitz-Gilbert (LLG) equation[60] and ensure the conservation of magnetic moments' magnitude[33], we implement the semi-implicit SIB method introduced by Mentink *et al.*[45] in LAMMPS and SPIN package. We use the molecular dynamics scheme (NVT or NPT) and SIB method to simulate the dynamics of spins and atomic motions. We control temperature using the Nosé Hoover thermostat with a temperature damping parameter of 100 time steps, and pressure using the Parrinello-Rahman barostat with a temperature damping parameter of 1000 time steps. To simulate the temperature-driven magnetic transition, we use NVT simulation to simulate the dynamics of atomic and spin motions with a damping coefficient in the LLG equation of 0.1. The magnetic moments of Cr are set as a constant at 3 $\mu_B$ during simulations.

For relaxation procedures of monolayer $CrI_3$, we use the CG and annealing algorithms. First, we optimize the initial random spin configuration using the CG algorithm with 1000 steps. Next, we run the NPT-SIB simulation with the annealing algorithm, and heat the system to 200 K in 10 ps. We then cool the system down to 0.01 K in 10 ps and equilibrate the systems for 5 ps. Finally, we relax the spin configuration with the CG algorithm in 10000 steps to reach the final relaxed magnetic structure.

We carry out simulations of $CrI_3$ for phase transition investigation on a periodic 24×24×1 supercell comprising 4608 atoms including 1202 Cr atoms using MD-SIB with a time step of 0.5 fs across 16 NVIDIA A800 GPUs simultaneously. At each given temperature, an equilibrium run of 5 ps is followed by a production run of 30 ps. A similar procedure is carried out on a periodic 30×16$\sqrt{3}$×1 monolayer $CrTe_2$ supercell comprising 2880 atoms including 960 Cr atoms across 8 NVIDIA A800 GPUs simultaneously in MD-SIB with a time step of 0.5 fs. At each given temperature, an equilibrium run of 5 ps is followed by a production run of 20 ps.

**Reference training sets**

For the dataset of monolayer $CrI_3$ and $CrTe_2$, we perform DFT calculations using VASP[61–63] with the Projector Augmented Wave (PAW) method[64] and the Perdew–Burke–Ernzerhof (PBE) functional[65]. We first generate atomic configurations through several NPT molecular dynamics simulations starting from 3×3×1 monolayer $CrI_3$ supercell with 72 atoms and 4×4×1 monolayer $CrTe_2$ supercell with 48 atoms. Then static spin-polarized calculations considering SOC[66] are carried out with randomly sampled spin configurations. We choose an energy cutoff for the plane wave basis of 400 eV and a single gamma-centered k-point grid for $CrI_3$ and a 2×3×1 Monkhorst-Pack k-point grid for $CrTe_2$. Total energy calculations are converged to $10^{-6}$ eV per supercell to ensure high-accuracy DFT data. For the localized electrons of Cr ions, we use a self-consistent value of the effective Hubbard U[67] of 4 eV for $CrI_3$. As for $CrTe_2$, the rotationally invariant DFT+U calculation[68] is carried with the U as 3 eV and J as 0.6 eV for Cr.

## Data availability

The authors declare that all data supporting the findings of this study are available from the corresponding author on reasonable request.

## Code availability

Time-reversal Euclidean neural networks package based on e3nn is available at https://github.com/Hongyu-yu/T-e3nn. Additional codes related to this work will be made available once the manuscript has been accepted for publication.

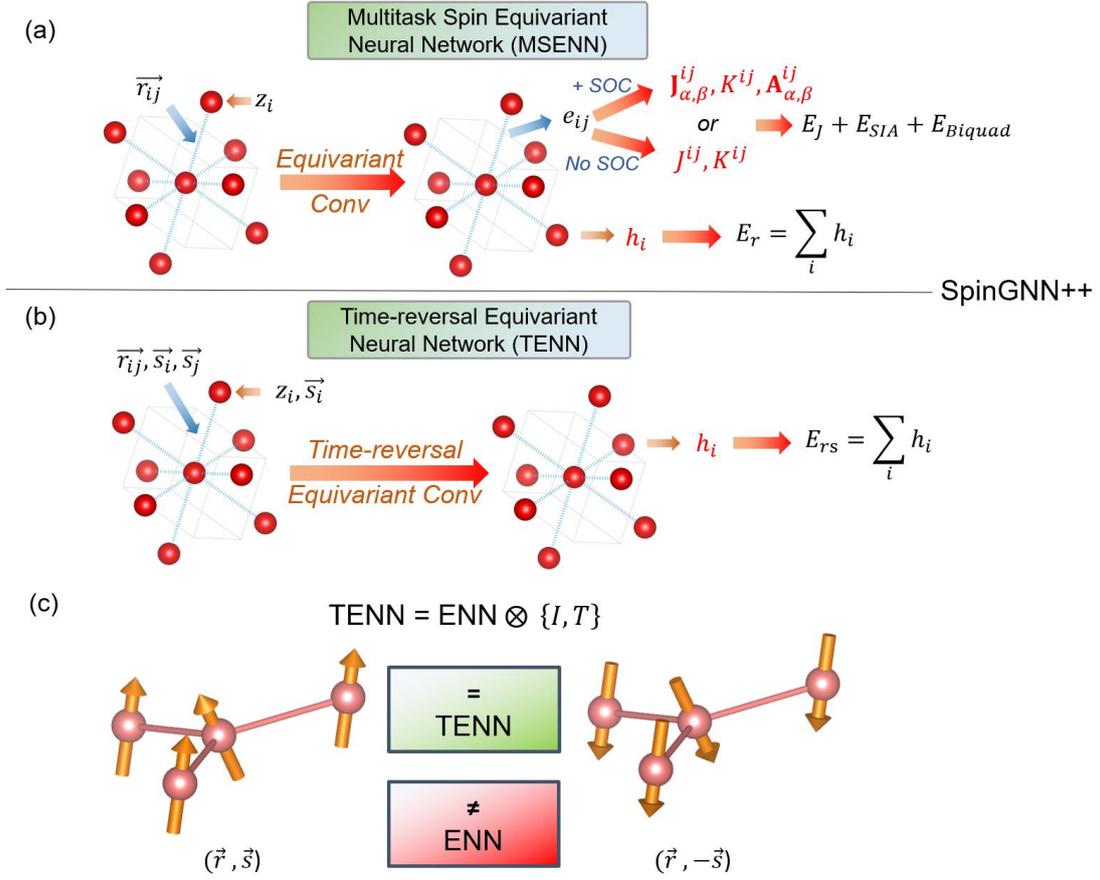

**Figure 1. The SpinGNN++ framework and illustration of time-reversal.** SpinGNN++ includes the multitask spin equivariant neural network (MSENN) and time-reversal equivariant neural network (TENN). **a** TENN utilizes the updated scalar and tensor features on edges of ENN to predict the J matrix, biquadratic coefficient, and single-ion anisotropy matrix if the spin-orbital coupling is considered. If not, the Heisenberg and biquadratic coefficients are predicted. The Heisenberg, Dzyaloshinskii–Moriya (DM), and Kitaev interactions are included in the J matrix. **b** TENN utilizes the time-reversal equivariant convolutions to build a high-order many-body general magnetic potential. **c** shows the atomic structure with spin configuration $(\vec{r}, \vec{s})$ and its configuration with a time-reversal operation $(\vec{r}, -\vec{s})$.

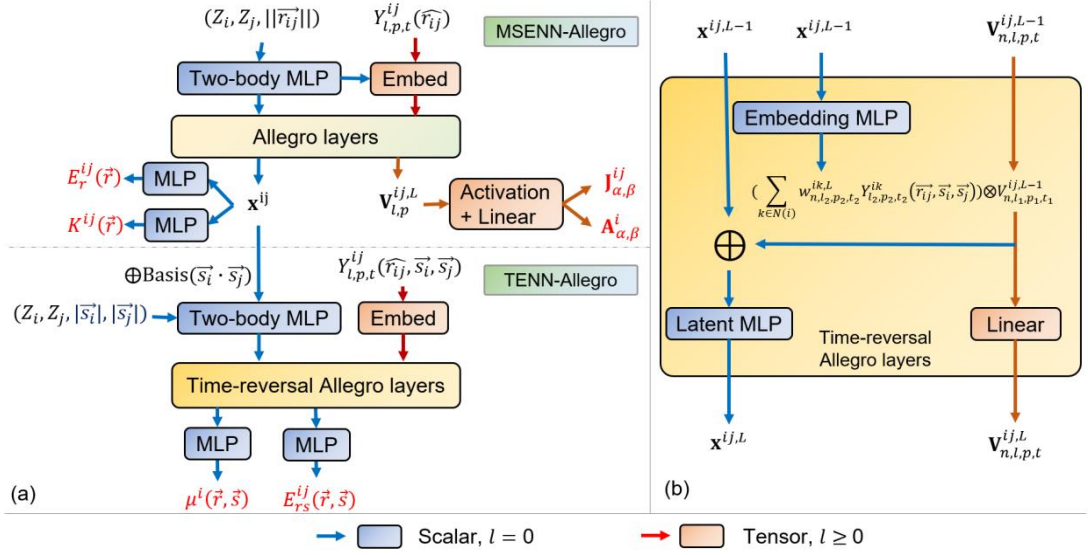

**Figure 2. The Spin-Allegro++ architecture. a** shows details of the Spin-Allegro++ and **b** details the time-reversal tensor product layer. Blue and red arrows and blocks represent scalar and tensor information and operation, respectively, $\otimes$ is the tensor product and $\oplus$ is concatenation. Output values are highlighted in red font. The three subscripts represent ($l$, $p$, $t$) for rotation order, parity, and time-reversal index and superscript for atom index.

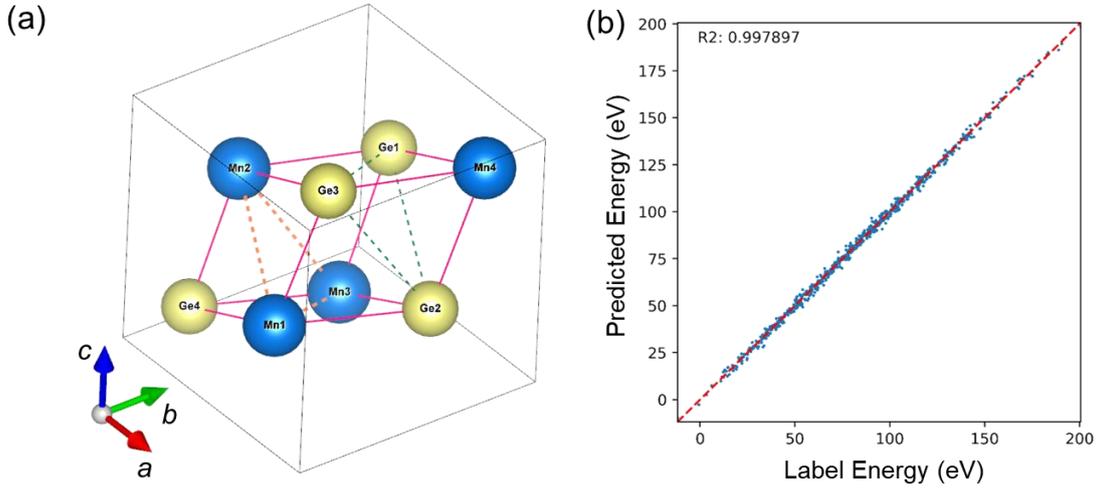

**Figure 3. MnGe-type structure and performance of Spin-Allegro++.** The first nearest neighbors between different atoms are connected by the lines.

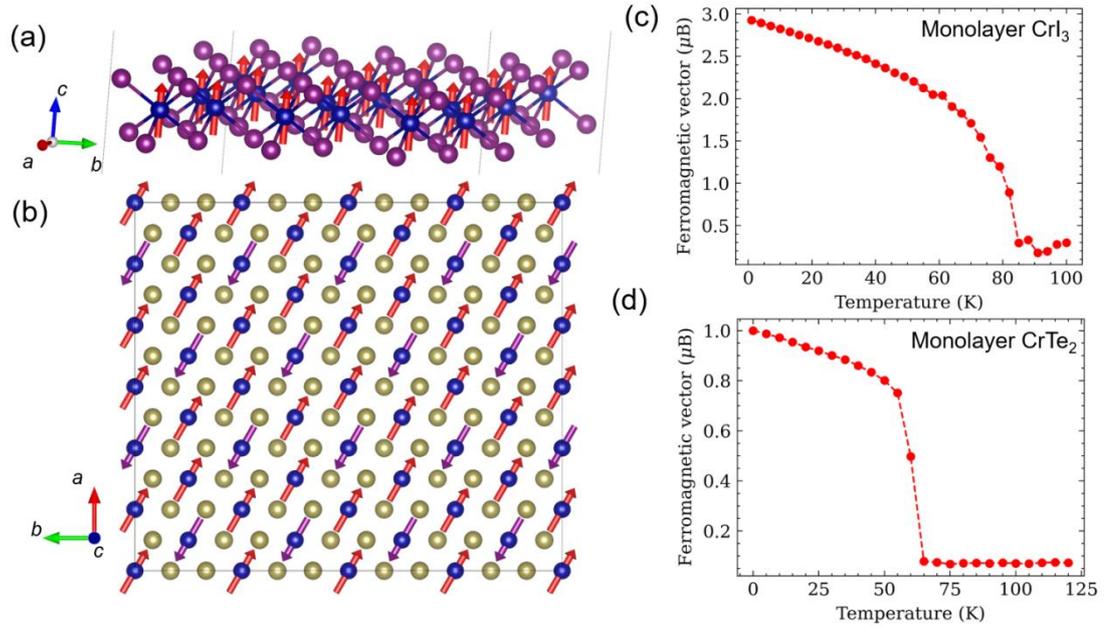

**Figure 4. Monolayer CrI$_3$ and CrTe$_2$ relaxed structure and phase transition.** **a** shows the relaxed ground structure of monolayer CrI$_3$ in the out-of-plane ferromagnetic state and **b** shows the relaxed ground structure of monolayer CrTe$_2$ with 1/3 of the Cr atoms exhibiting an "up" orientation in purple vectors surrounding by 2/3 displaying a "down" in red vectors. **c** and **d** shows the system average ferromagnetic vector of Cr atoms in different temperatures of monolayer CrI$_3$ and CrTe$_2$ from simulations of spin-lattice dynamics with Spin-Allegro++. Obvious magnetic transition is observed and transition temperature is estimated to be about 70 K and 60 K, respectively for monolayer CrI$_3$ and CrTe$_2$.

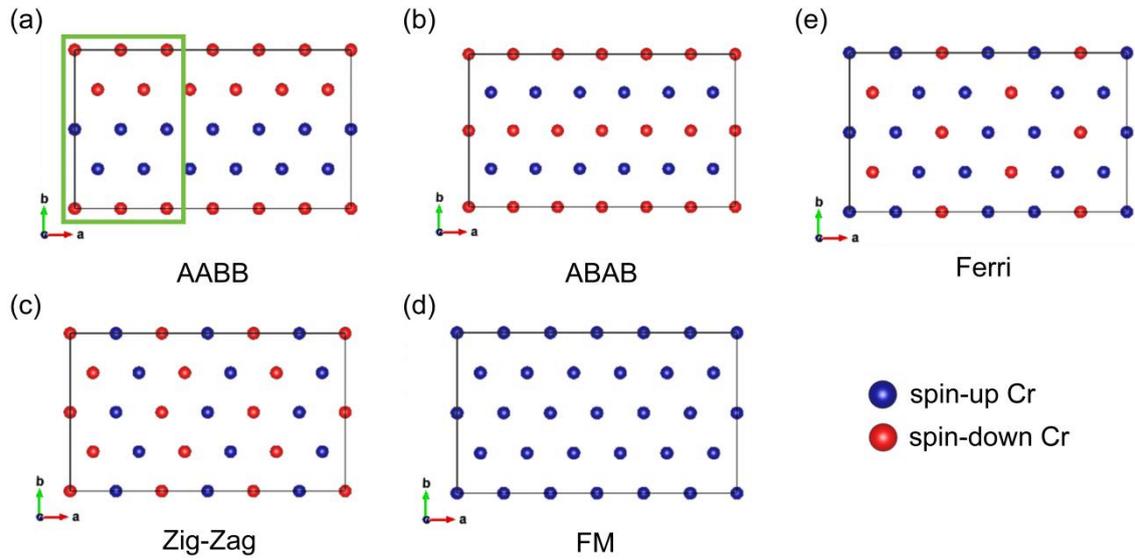

**Figure 5. Magnetic configurations and cell adopted in the phase diagram calculation of monolayer CrTe$_2$.** Only Cr atoms are shown with up and down orientation in different colors. The commonly used cell in monolayer CrTe$_2$ calculations is illustrated by the green rectangle in (a), which is unable to accommodate the Ferri state.

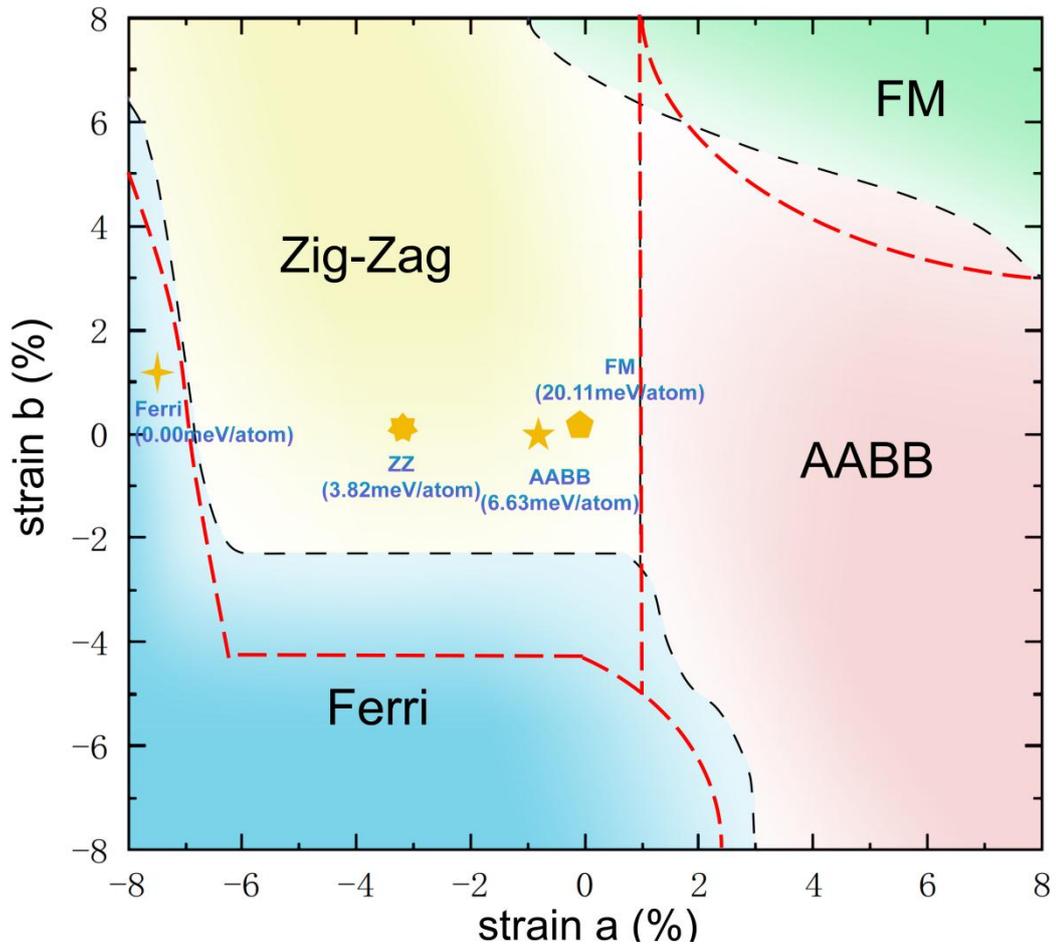

**Figure 6. The magnetic phase diagram of monolayer CrTe$_2$.** The black and red dashed lines define the boundaries for the cases without and with consideration of spin-orbit coupling, respectively. Meanwhile, the relaxed freestanding structures of zig-zag, ferrimagnetic (Ferri), AABB, and ferromagnetic (FM) are represented by yellow symbols. Ferri is the magnetic state with the lowest global energy, and the energy differences between Ferri and other states are indicated by the yellow symbols.

# Reference

1. Schleder, G. R., Padilha, A. C. M., Acosta, C. M., Costa, M. & Fazzio, A. From DFT to machine learning: recent approaches to materials science–a review. *J. Phys. Mater.* **2**, 032001 (2019).
2. Carleo, G. *et al.* Machine learning and the physical sciences. *Rev. Mod. Phys.* **91**, 045002 (2019).
3. Agrawal, A. & Choudhary, A. Perspective: Materials informatics and big data: Realization of the "fourth paradigm" of science in materials science. *APL Mater.* **4**, 053208 (2016).
4. Hu, L., Huang, B. & Liu, F. Atomistic Mechanism Underlying the Si(111)-(7×7) Surface Reconstruction Revealed by Artificial Neural-Network Potential. *Phys. Rev. Lett.* **126**, 176101 (2021).
5. Musaelian, A. *et al.* Learning local equivariant representations for large-scale atomistic dynamics. *Nat. Commun.* **14**, 579 (2023).
6. Himanen, L. *et al.* DScribe: Library of descriptors for machine learning in materials science. *Comput. Phys. Commun.* **247**, 106949 (2020).
7. Isayev, O. *et al.* Universal fragment descriptors for predicting properties of inorganic crystals. *Nat. Commun.* **8**, 15679 (2017).
8. Smith, J. S., Isayev, O. & Roitberg, A. E. ANI-1: an extensible neural network potential with DFT accuracy at force field computational cost. *Chem. Sci.* **8**, 3192–3203 (2017).
9. Behler, J. Atom-centered symmetry functions for constructing high-dimensional neural network potentials. *J. Chem. Phys.* **134**, 074106 (2011).
10. Schütt, K. T., Sauceda, H. E., Kindermans, P.-J., Tkatchenko, A. & Müller, K.-R. SchNet – A deep learning architecture for molecules and materials. *J. Chem. Phys.* **148**, 241722 (2018).
11. Xie, T. & Grossman, J. C. Crystal Graph Convolutional Neural Networks for an Accurate and Interpretable Prediction of Material Properties. *Phys. Rev. Lett.* **120**, 145301 (2018).
12. Chen, C., Ye, W., Zuo, Y., Zheng, C. & Ong, S. P. Graph Networks as a Universal Machine Learning Framework for Molecules and Crystals. *Chem. Mater.* **31**, 3564–3572 (2019).
13. Gasteiger, J., Groß, J. & Günnemann, S. Directional Message Passing for Molecular Graphs. Preprint at https://doi.org/10.48550/arXiv.2003.03123 (2022).
14. Schmidt, J., Pettersson, L., Verdozzi, C., Botti, S. & Marques, M. A. L. Crystal graph attention networks for the prediction of stable materials. *Sci. Adv.* **7**, eabi7948 (2021).
15. Choudhary, K. & DeCost, B. Atomistic Line Graph Neural Network for improved materials property predictions. *Npj Comput. Mater.* **7**, 1–8 (2021).
16. Chen, C. & Ong, S. P. A universal graph deep learning interatomic potential for the periodic table. *Nat. Comput. Sci.* **2**, 718–728 (2022).
17. Deng, B. *et al.* CHGNet as a pretrained universal neural network potential for charge-informed atomistic modelling. *Nat. Mach. Intell.* **5**, 1031–1041 (2023).

18. Kondor, R., Lin, Z. & Trivedi, S. Clebsch– Gordan Nets: a Fully Fourier Space Spherical Convolutional Neural Network. in *Advances in Neural Information Processing Systems* vol. 31 (Curran Associates, Inc., 2018).
19. Weiler, M., Geiger, M., Welling, M., Boomsma, W. & Cohen, T. S. 3D Steerable CNNs: Learning Rotationally Equivariant Features in Volumetric Data. in *Advances in Neural Information Processing Systems* vol. 31 (Curran Associates, Inc., 2018).
20. Thomas, N. *et al.* Tensor field networks: Rotation- and translation-equivariant neural networks for 3D point clouds. Preprint at https://doi.org/10.48550/arXiv.1802.08219 (2018).
21. Kondor, R. N-body Networks: a Covariant Hierarchical Neural Network Architecture for Learning Atomic Potentials. Preprint at https://doi.org/10.48550/arXiv.1803.01588 (2018).
22. Batzner, S. *et al.* E(3)-equivariant graph neural networks for data-efficient and accurate interatomic potentials. *Nat. Commun.* **13**, 2453 (2022).
23. Batatia, I., Kovacs, D. P., Simm, G., Ortner, C. & Csanyi, G. MACE: Higher Order Equivariant Message Passing Neural Networks for Fast and Accurate Force Fields. *Adv. Neural Inf. Process. Syst.* **35**, 11423–11436 (2022).
24. Yu, H. *et al.* Complex spin Hamiltonian represented by an artificial neural network. *Phys. Rev. B* **105**, 174422 (2022).
25. Yu, H. *et al.* Spin-Dependent Graph Neural Network Potential for Magnetic Materials. Preprint at https://doi.org/10.48550/arXiv.2203.02853 (2023).
26. Eckhoff, M. & Behler, J. High-dimensional neural network potentials for magnetic systems using spin-dependent atom-centered symmetry functions. *Npj Comput. Mater.* **7**, 170 (2021).
27. Nikolov, S. *et al.* Data-driven magneto-elastic predictions with scalable classical spin-lattice dynamics. *Npj Comput. Mater.* **7**, 153 (2021).
28. Chapman, J. B. J. & Ma, P.-W. A machine-learned spin-lattice potential for dynamic simulations of defective magnetic iron. *Sci. Rep.* **12**, 22451 (2022).
29. Domina, M., Cobelli, M. & Sanvito, S. Spectral neighbor representation for vector fields: Machine learning potentials including spin. *Phys. Rev. B* **105**, 214439 (2022).
30. Novikov, I., Grabowski, B., Körmann, F. & Shapeev, A. Magnetic Moment Tensor Potentials for collinear spin-polarized materials reproduce different magnetic states of bcc Fe. *Npj Comput. Mater.* **8**, 13 (2022).
31. Rinaldi, M., Mrovec, M., Bochkarev, A., Lysogorskiy, Y. & Drautz, R. Non-collinear Magnetic Atomic Cluster Expansion for Iron. *arXiv.org* https://arxiv.org/abs/2305.15137v1 (2023).
32. Neaton, J. B., Ederer, C., Waghmare, U. V., Spaldin, N. A. & Rabe, K. M. First-principles study of spontaneous polarization in multiferroic BiFeO3. *Phys. Rev. B* **71**, 014113 (2005).
33. Wang, D., Weerasinghe, J. & Bellaiche, L. Atomistic Molecular Dynamic Simulations of Multiferroics. *Phys. Rev. Lett.* **109**, 067203 (2012).
34. Coldea, R. *et al.* Spin Waves and Electronic Interactions in LaCuO4. *Phys. Rev.*

*Lett.* **86**, 5377–5380 (2001).

35. Geiger, M. & Smidt, T. e3nn: Euclidean Neural Networks. (2022) doi:10.48550/arXiv.2207.09453.
36. Takizawa, H., Sato, T., Endo, T. & Shimada, M. High-pressure synthesis and electrical and magnetic properties of MnGe and CoGe with the cubic B20 structure. *J. Solid State Chem.* **73**, 40–46 (1988).
37. Lou, F. *et al.* PASP: Property analysis and simulation package for materials. *J. Chem. Phys.* **154**, 114103 (2021).
38. Huang, B. *et al.* Layer-dependent ferromagnetism in a van der Waals crystal down to the monolayer limit. *Nature* **546**, 270–273 (2017).
39. Dillon, J. F., Jr. & Olson, C. E. Magnetization, Resonance, and Optical Properties of the Ferromagnet CrI3. *J. Appl. Phys.* **36**, 1259–1260 (2004).
40. Lado, J. L. & Fernández-Rossier, J. On the origin of magnetic anisotropy in two dimensional CrI3. *2D Mater.* **4**, 035002 (2017).
41. Xu, C., Feng, J., Xiang, H. & Bellaiche, L. Interplay between Kitaev interaction and single ion anisotropy in ferromagnetic CrI3 and CrGeTe3 monolayers. *Npj Comput. Mater.* **4**, 57 (2018).
42. Xu, C. *et al.* Possible Kitaev Quantum Spin Liquid State in 2D Materials with S=3/2. *Phys. Rev. Lett.* **124**, 087205 (2020).
43. Hestenes, M. R. & Stiefel, E. Methods of conjugate gradients for solving linear systems. *J. Res. Natl. Bur. Stand.* **49**, 409 (1952).
44. Kirkpatrick, S., Gelatt, C. D. & Vecchi, M. P. Optimization by Simulated Annealing. *Science* **220**, 671–680 (1983).
45. Mentink, J. H., Tretyakov, M. V., Fasolino, A., Katsnelson, M. I. & Rasing, T. Stable and fast semi-implicit integration of the stochastic Landau–Lifshitz equation. *J. Phys. Condens. Matter* **22**, 176001 (2010).
46. Sun, X. *et al.* Room temperature ferromagnetism in ultra-thin van der Waals crystals of 1T-CrTe2. *Nano Res.* **13**, 3358–3363 (2020).
47. Zhang, X. *et al.* Room-temperature intrinsic ferromagnetism in epitaxial CrTe2 ultrathin films. *Nat. Commun.* **12**, 2492 (2021).
48. Xian, J.-J. *et al.* Spin mapping of intralayer antiferromagnetism and field-induced spin reorientation in monolayer CrTe2. *Nat. Commun.* **13**, 257 (2022).
49. Lv, H. Y., Lu, W. J., Shao, D. F., Liu, Y. & Sun, Y. P. Strain-controlled switch between ferromagnetism and antiferromagnetism in 1T−CrX2(X=Se,Te) monolayers. *Phys. Rev. B* **92**, 214419 (2015).
50. Gao, P., Li, X. & Yang, J. Thickness Dependent Magnetic Transition in Few Layer 1T Phase CrTe2. *J. Phys. Chem. Lett.* **12**, 6847–6851 (2021).
51. Liu, Y., Kwon, S., de Coster, G. J., Lake, R. K. & Neupane, M. R. Structural, electronic, and magnetic properties of CrTe2. *Phys. Rev. Mater.* **6**, 084004 (2022).
52. Wu, L., Zhou, L., Zhou, X., Wang, C. & Ji, W. In-plane epitaxy-strain-tuning intralayer and interlayer magnetic coupling in CrSe2 and CrTe2 monolayers and bilayers. *Phys. Rev. B* **106**, L081401 (2022).
53. Zhu, H., Gao, Y., Hou, Y., Gui, Z. & Huang, L. Insight into strain and electronic correlation dependent magnetism in monolayer 1T−CrTe2. *Phys. Rev. B* **108**,

144404 (2023).

54. Musaelian, A., Johansson, A., Batzner, S. & Kozinsky, B. Scaling the leading accuracy of deep equivariant models to biomolecular simulations of realistic size. Preprint at https://doi.org/10.48550/arXiv.2304.10061 (2023).
55. Paszke, A. *et al.* PyTorch: An Imperative Style, High-Performance Deep Learning Library. in *Advances in Neural Information Processing Systems* vol. 32 (Curran Associates, Inc., 2019).
56. Momma, K. & Izumi, F. VESTA: a three-dimensional visualization system for electronic and structural analysis. *J. Appl. Crystallogr.* **41**, 653–658 (2008).
57. Hunter, J. D. Matplotlib: A 2D Graphics Environment. *Comput. Sci. Eng.* **9**, 90–95 (2007).
58. Tranchida, J., Plimpton, S. J., Thibaudeau, P. & Thompson, A. P. Massively parallel symplectic algorithm for coupled magnetic spin dynamics and molecular dynamics. *J. Comput. Phys.* **372**, 406–425 (2018).
59. Thompson, A. P. *et al.* LAMMPS - a flexible simulation tool for particle-based materials modeling at the atomic, meso, and continuum scales. *Comput. Phys. Commun.* **271**, 108171 (2022).
60. García-Palacios, J. L. & Lázaro, F. J. Langevin-dynamics study of the dynamical properties of small magnetic particles. *Phys. Rev. B* **58**, 14937–14958 (1998).
61. Kresse, G. & Hafner, J. Ab initio molecular dynamics for liquid metals. *Phys. Rev. B* **47**, 558 (1993).
62. Kresse, G. & Hafner, J. Ab initio molecular-dynamics simulation of the liquid-metal--amorphous-semiconductor transition in germanium. *Phys. Rev. B* **49**, 14251–14269 (1994).
63. Kresse, G. & Furthmüller, J. Efficiency of ab-initio total energy calculations for metals and semiconductors using a plane-wave basis set. *Comput. Mater. Sci.* **6**, 15–50 (1996).
64. Blöchl, P. E. Projector augmented-wave method. *Phys. Rev. B* **50**, 17953 (1994).
65. Perdew, J. P., Burke, K. & Ernzerhof, M. Generalized Gradient Approximation Made Simple. *Phys. Rev. Lett.* **77**, 3865–3868 (1996).
66. Ma, P.-W. & Dudarev, S. L. Constrained density functional for noncollinear magnetism. *Phys. Rev. B* **91**, 054420 (2015).
67. Dudarev, S. L., Botton, G. A., Savrasov, S. Y., Humphreys, C. J. & Sutton, A. P. Electron-energy-loss spectra and the structural stability of nickel oxide: An LSDA+U study. *Phys. Rev. B* **57**, 1505–1509 (1998).
68. Liechtenstein, A. I., Anisimov, V. I. & Zaanen, J. Density-functional theory and strong interactions: Orbital ordering in Mott-Hubbard insulators. *Phys. Rev. B* **52**, R5467–R5470 (1995).

## Acknowledgments

We acknowledge financial support from the National Key R&D Program of China (No. 2022YFA1402901), NSFC (No. 11825403, 11991061, 12188101, 12174060, and 12274082), the Guangdong Major Project of the Basic and Applied Basic Research (Future functional materials under extreme conditions--2021B0301030005),

and Shanghai Pilot Program for Basic Research—Fudan University 21TQ1400100

(23TQ017). C.X. also acknowledges support from the Shanghai Science and Technology Committee (Grant No. 23ZR1406600).

## Author contributions

H.Y.Y. conceived the model architecture, derived the theoretical analysis, implemented the software, generated the datasets, trained models, ran simulations, analysed the results, and wrote the first version of the manuscript. B.Y.L. generated the $CrTe_2$ dataset, carried out related first principles calculations of $CrTe_2$, analysed the results and wrote related content of the first version of the manuscript. Y.Z. contributed to the discussion of the method implementation. J.Y.J contributed to the discussion of the theoretical analysis. L.H.H contributed to the spin dynamics implementation in LAMMPS. C.S.X and X.G.G. contributed to the discussion of the methods and results. H.J.X. supervised and guided the project from conception to design of experiments, methods, theory, and implementation, as well as analysis of data and results. All authors contributed to the manuscript.

## Competing interests

The authors declare no competing interests.

## Additional information

**Correspondence** and requests for materials should be addressed to Hongjun Xiang.